\documentclass[fleqn,12pt]{wlscirep}

\usepackage[utf8]{inputenc}
\usepackage[T1]{fontenc}

\usepackage{hyperref}
\usepackage{xr-hyper}
\usepackage{times}
\usepackage{epsfig}
\usepackage{graphicx}
\usepackage{amsmath}
\usepackage{amssymb}
\usepackage{soul,xcolor}        
\usepackage{dirtytalk}          
\usepackage{soul}               
\usepackage{url}

\usepackage{booktabs} 
\usepackage{tabularx}
\usepackage{multirow}
\usepackage{adjustbox} 
\usepackage{tabu}
\usepackage{float}
\usepackage{siunitx}
\usepackage[table]{xcolor}

\newcolumntype{Y}{>{\centering\arraybackslash}X}

\title{Beyond Localisation Accuracy: Sensorimotor Effects of HRTF Individualisation}

\author[1,*]{Fulvio Missoni}
\author[2]{Katarina C. Poole}
\author[2]{Tim Murray-Browne}
\author[1,+]{Andrea Canessa}
\author[2,+]{Lorenzo Picinali}

\affil[1]{University of Genoa, DIBRIS, Genoa, 16145, Italy}
\affil[2]{Dyson School of
Design Engineering, Imperial College London, London, United Kingdom}

\affil[*]{fulvio.missoni@edu.unige.it}

\affil[+]{these authors contributed equally to this work}


\begin{abstract}
Everyday spatial listening requires the brain to integrate cues from the body, environment, other senses, and movement, continuously translating auditory information into action. Yet HRTF individualisation is still commonly assessed through localisation accuracy, which may not fully capture its effects on this sensorimotor process. Here, we investigate whether these effects can instead be revealed through behaviour in a more ecologically valid listening task.

We used an aurally guided visual search paradigm in which listeners located a visual target using a co-located virtual sound while moving freely, comparing individualised and non-individualised HRTFs under anechoic and reverberant conditions. Performance was assessed through response times and measures of movement organisation. In anechoic conditions, individualised HRTFs produced faster responses than non-individualised HRTFs, with an average reduction of approximately 200 ms and the clearest benefit for front–back source locations. This advantage was expressed primarily in movement initiation, whereas overall movement extent was only weakly affected. Under reverberant conditions, HRTF-dependent differences disappeared.

These results suggest that HRTF individualisation can influence how listeners plan and initiate orienting actions even when differences in conventional localisation outcomes are limited. Assessing sensorimotor behaviour alongside localisation performance may therefore provide a more sensitive and ecologically relevant account of the perceptual benefits of HRTF individualisation.

\end{abstract}

\begin{document}

\flushbottom
\maketitle

\thispagestyle{empty}

\section*{Introduction}
    A large part of what we know about spatial hearing comes from static, blindfolded localisation tasks performed under acoustically controlled conditions \cite{carliniAuditoryLocalizationComprehensive2024a, blauert1997spatial}. These constraints isolate auditory processing from natural multimodal interactions, such as vision and active head movement \cite{wallachRoleHeadMovements1940} and have been essential to establish the foundational role of the head-related transfer function (HRTF) in spatial hearing. By capturing the direction-dependent filtering imposed by the head, torso and pinnae \cite{blauert1997spatial}, the HRTF encodes spatial information into distinct auditory cues. These can be separated into binaural cues (i.e., interaural time and level differences) for lateral localisation, and monaural spectral cues that resolve elevation and front-back confusion.
Since morphology of external-ear varies substantially across individuals \cite{algazi2001cipic}, whether \cite{geronazzo2025strong} and how \cite{fantini2025survey} to individualise HRTFs has become a central question in spatial hearing research. Static localisation tasks have served as the primary benchmark to evaluate it, confirming a consistent group-level benefit \cite{mendoncaReviewAuditorySpace2014a}. At the individual level, however, behavioural outcomes become unreliable \cite{daugintis2025comparing-key, andreopoulou2016investigation, kim2020investigation} as non-acoustic factors, such as listener task familiarity and perceptual sensitivity, account for more outcome variance than the HRTF itself \cite{zagala2020comparison, majdak2014acoustic}.

To overcome the artificial nature of static tasks, recent research has increasingly directed its efforts toward more realistic listening contexts (e.g., allowing free head movement, embedding complex auditory tasks such as speech-in-noise, or relying on qualitative perceptual judgements). Across these approaches, however, results have been inconsistent. Free head movement, for instance, introduces dynamic variation in binaural cues that helps resolve directional ambiguities, reducing reliance on spectral cues that HRTFs provide and masking their specific contribution \cite{begaultDirectComparisonImpact2001, oberemExperimentsLocalizationAccuracy2020, rummukainenHeadRelatedTransferFunctions2021}. Yet when movement is restricted to a limited range ($\leq 30^\circ$), HRTF individualisation has been shown to remain beneficial \cite{ben-hurLocalizationVirtualSounds2020}. Speech-in-noise tasks reveal an individual-HRTF advantage only when listeners are prevented from relying on non-spatial cues \cite{gonzalez2024spatial, vicente2025exploring}. Finally, qualitative plausibility ratings show a high tolerance for non-individualised audio, with no significant difference in perceived quality between individual and non-individual HRTFs in anechoic \cite{lubeck2023investigation} or reverberant environments \cite{blauRealisticBinauralAuralizations2021}, or even within interactive, audiovisual VR scenes \cite{starz2025comparison}. This has been attributed to an imprecise internal reference against which listeners judge plausibility \cite{neidhardt2021availability}. 

Rather than concluding that listeners are simply perceptually insensitive to individualisation, a more fundamental diagnosis is that these distinct paradigms often still isolate hearing from the natural, continuous coupling between auditory perception and physical action \cite{froese2020passive}. From a sensorimotor perspective, changes to auditory spatial cues can consequently lead listeners to adapt how they relate perceived sound locations to their actions \cite{vanOpstal2016, hofmanRelearningSound1998, vanWanrooijRelearningSoundLocalization2005}. This adaptation does not necessarily result in changes in localisation accuracy, and may instead be reflected in changes in how listeners move or orient during listening \cite{valzolgherAdaptingAlteredAuditory2022}. Considering movement alongside accuracy may therefore provide a broader view of how listeners adapt to changes in auditory spatial cues. When experimental paradigms capture this full complexity—allowing listeners to move freely and act within a coherent, multisensory context—the benefits of individualisation emerge more consistently. For example, Jenny et al. (2020) \cite{jennyUsabilityIndividualizedHeadRelated2020} demonstrated that in an interactive VR scene, where free movement and visual realism are intrinsically linked, individual HRTFs significantly enhanced both localisation behaviour and perceived realism. This highlights that combining active movement with a visual context is key to capturing the true impact of individualised spatial cues.

The aurally-guided visual search task, introduced over three decades ago by Perrott et al. (1990) \cite{perrottAuditoryPsychomotorCoordination1990} and refined by Bolia et al. (1999) \cite{boliaAurallyAidedVisual1999}, provides an ideal framework to achieve this balance of multisensory complexity, natural motor action, and strict experimental control. In this paradigm, listeners are required to locate a single visual target hidden among numerous visual distractors, guided purely by a co-located sound. Its high ecological validity stems from the fact that it elicits natural, unconstrained orienting behaviour and directly uses that behaviour as the primary metric of performance. Reliable spatial hearing allows the listener to rapidly map the sound to a physical location and orient toward it, resulting in a short search time with highly efficient head and eye exploration. Previous work has successfully linked longer search times and altered movement strategies in this paradigm to front–back confusion under altered spectral cues \cite{lladoImpactHeadwornDevices2024} and degraded cues from hearing protectors \cite{simpsonImpactHearingProtection2005,simpsonAURALLYAIDEDVISUAL2010}. Furthermore, it remains highly practical for use within a short assessment window.

In this study, we apply the aurally-guided visual search paradigm to evaluate HRTF individualisation for the first time. Rather than relying on static pointing or subjective ratings, we use this ecologically valid framework to determine whether individual HRTFs yield measurable benefits in natural movement strategies and response times. Specifically, we investigate whether the task is sensitive to HRTF individualisation within a short assessment window, how these sensorimotor benefits are modulated by the acoustic environment (comparing anechoic rendering to conditions with early reflections), and whether individualisation influences movement planning independently of overall accuracy.
Based on the reviewed literature, we hypothesise:
\begin{itemize}
\item [H1.] Individualised HRTFs will produce stronger, more consistent behavioural benefits in anechoic conditions.

\item [H2.] HRTF individualisation will alter listeners' movement search strategies and planning.

\item [H3.] Early reflections will attenuate or eliminate the advantage of individualised HRTFs, as room acoustics reduce the spectral detail on which individualisation relies.
\end{itemize}

\section*{Methods}
\paragraph{Participants}
12 self-reported normal hearing participants aged between 20 and 48 years were recruited for this study (2 females, 10 males; age: mean = 28.75, SD = 7). 
Participants were tested under four conditions where different auditory alterations were used. In particular, we used individual and non-individual HRTFs, combined with two different levels of reverberation: anechoic (i.e., no reverberation) and reverberation of a small room (width 3 m, depth 4 m, height 2.5 m).
This study was approved by the Imperial College Research Ethics Committee and performed following relevant guidelines and regulations. Participants signed a consent form to take part in the experiment, in which they were informed about the procedure and the purpose of the study, and that they may withdraw at any time.

\paragraph{Equipment and experimental setup}
The experiment was conducted in virtual reality using a custom Android app, developed in Unity (version 2022.3.17f1).
The auditory and the visual parts of the virtual stimuli were presented respectively, through a pair of over-ear headphones (Sennheiser HD599SE) and a head-mounted display (HMD) (Meta Quest 3; Meta Corporation, USA). This standalone HMD operates on a 64-bit Android operating system and is equipped with: a Qualcomm Snapdragon XR2 Gen2, 8GB RAM (LPDDR5), an Adreno 740 graphics processor, and 512 GB storage. Participants could interact inside the virtual scene using the two handheld Meta controllers. 
The VR system was tracked in real-time with inside-out tracking in six degrees of freedom (DoF), with a sample rate of 72Hz.
The acoustic scenes were rendered in real-time using the Unity Wrapper module of the 3DTune-In Toolkit \cite{cuevas-rodriguez3DTuneInToolkit2019}, using the head tracking data of the HMD.

\paragraph{Stimuli}
The stimulus was a white noise burst (calibrated level: 70dBA at a distance of 1.8 m from the listener’s head position)) lasting 12 seconds in order to allow participants to use auditory dynamic cues during the entire duration of a single trial. The stimulus was a repetition of an amplitude-modulated white-noise signal (total duration: 300ms, modulation frequency: 5Hz), followed by 0.5s of silence. 
Stimuli were rendered using either individual or non-individual HRTFs, depending on the tested condition. 
HRTFs were measured following the the exact measurement protocol, hardware configuration, and signal processing steps detailed in \cite{engelSONICOMHRTFDataset2023}. For the purpose of this study, we utilised individual HRTFs of each participant and a non-individual HRTF measured on a KEMAR mannequin to provide a baseline for comparison.
Two different reverberation conditions were considered. The first was an anechoic condition, in which only the direct sound was reproduced. The second was a reverberant condition, implemented using the Reverberant Virtual Loudspeaker (RVL) method \cite{engelReverberationItsBinaural2022}.

For the reverberant condition, we used room impulse responses measured in a small trapezoidal meeting room described in \cite{engel2021perceptual} This room corresponds to a typical moderately reverberant indoor space, with a carpeted floor, slightly asymmetrical walls, and limited furnishing. Its acoustic characteristics provide a realistic yet controlled reverberant environment suitable for binaural rendering. Table \ref{table:room_reflections} summarizes the main parameters of the room reflections.


\begin{table}[h!]
\centering
 \begin{tabular}{ c | c | c | c| c | c | c } 
 \hline
   Frq.band [Hz] & 250 & 500 & 1000 & 2000 & 4000 & 8000\\
 \hline\hline

 RT [s] & 0.78 & 0.63 & 0.53 & 0.48 & 0.49 & 0.46\\ 
 EDT [s] & 0.70 & 0.49 & 0.43 & 0.36 & 0.34 & 0.28\\ 
 \hline
 \end{tabular}
\caption{\textbf{Room reflections parameters.} Reverberation time (RT) and early decay time (EDT) for octave frequency bands (in Hz).}
\label{table:room_reflections}
\end{table}
The distance attenuation is a rather traditional approach \cite{chowningSimulationMovingSound1977}. The computation was independent for the direct ($-6$dB every doubling of distance) and reverberant ($-3$dB every doubling of distance) sound path to emulate what happens in real conditions \cite{cuevas-rodriguez3DTuneInToolkit2019}. All stimuli were generated and stored in $48$kHz, $16$-bit format.

\paragraph{Experimental procedure}\label{par:exp_proc}
The VR listening test took place in the same room used for acoustic measurements, which is a semi-anechoic room (width = 5m, length = 5m, height = 7.5m, background noise = 23 dB A-weighted). Participants were seated at the center of the room on an office chair that allowed free rotation while restricting translational movement, and wore a head-mounted display and headphones. In the virtual environment, listeners were surrounded by 28 spherically-organised virtual loudspeakers, each equipped with 4 LEDs on the frontal surface. Participants were instructed to locate a played target loudspeaker and report the number of LEDs on it by pressing the corresponding button of the controller as soon as they locate it. 
The number of lit LEDs distinguished target loudspeakers from distractors, with even numbers indicating targets and odd numbers representing distractors. To reduce the reliability of the visual modality, all the loudspeakers were used as distractors with 3 LEDs on, except for the played target.
At the start of each trial, participants were required to align their head-gaze with the frontal loudspeaker by superimposing a virtual cross with the central visual marker for 1 s to initiate the trial.
The loudspeakers were at 1.8 m, arranged as follows: 12 equally spaced in the horizontal plane (azimuth step: 30$^\circ$, elevation equal to 0$^\circ$), 8 in the upper plane (azimuth step: 45$^\circ$, elevation equal to 30$^\circ$), and 8 in the lower plane (azimuth step: 45$^\circ$, elevation equal to -30$^\circ$). 
This task was repeated under four different acoustic conditions obtained by the combinations of different HRTF (individual and non-individual) and reverberation levels (anechoic and early reverberation). Each location was repeated three times, for a total of 2 HRTFs $\times$ 2 reverberant levels $\times$ 28 positions $\times$ 3 repetitions = 336 trials collected from each participant.
The order of blocks and stimulus locations was randomised between subjects to minimise the impact of learning effects on the acoustic condition.

Before the testing phase, participants took part in a familiarisation block in which they performed the task with received feedback on the correctness of their responses after each trial via a virtual menu. The familiarisation was conducted with the individual HRTF in anechoic conditions.
Because the presence of visual feedback could promote rapid adaptation and learning of individual auditory cues, the familiarisation phase was deliberately kept brief. To this end, each sound source was presented only once per location, and stimulation was restricted to a subset of the loudspeaker positions. In particular, for each elevation (i.e., $30, -30$, and $0$ degrees), only lateral (azimuths of $90^\circ$ and $270^\circ$) and frontal positions (azimuths of $0^\circ$ and $180^\circ$) were considered. 

\paragraph{Behavioural and head movements metrics}
Task performance was assessed based on the number of correct responses and response time (RT), defined as the time interval between the onset of sound playback and the moment the button was pressed. To analyse listening strategies during task execution, head-tracking data (i.e., HMD rotation, sampled at 72 Hz) were recorded. Trials corresponding to the frontal loudspeaker at the horizontal plane were excluded from the analysis to avoid ceiling effects.

RT estimation was further refined to exclude the final portion of the recorded head-movement trajectory. This adjustment was motivated by observations that, in a small subset of trials, participants initiated a return movement toward the starting position before responding, artificially inflating RT estimates. To account for this behaviour, the following algorithm was applied.
First, the angular velocity was computed, using a Savitzsky–Golay finite impulse response filter to remove noise. A speed threshold is then applied to define the head movement in the opposite direction. The threshold was set at $5\%$ of the peak velocity. To obtain the response time of a head movement, the algorithm walked forward in time from the last velocity peak over the threshold and marked the response time as the last point where the head’s angular velocity was above the threshold. 

For the analysis of the listening strategy, head rotations were decomposed into two angular components using the truncated Fick-Gimbal reference system \cite{mclachlanHeadRotationsFollow2024}. In this system, rotations are represented as a sequence of horizontal rotation (\textit{yaw}) around the vertical axis and angular movement around the interaural axis (\textit{tilt}). Then, the obtained trial-by-trial trajectories were described in terms of the latency of movement initiation (or Movement Onset, MO) and the quantity of movement needed to identify the target, namely the Total Angular Extent (TAE).
Specifically, MO was determined moving backwards from the first velocity peak to identify the last time sample where yaw velocity remained above the threshold, set as the 5\% of the peak velocity. Since movements were mainly horizontal, we decided to estimate MO considering only yaw rotations. 
We decided to assess TAE as proposed by \cite{simpsonImpactHearingProtection2005} in a similar experimental paradigm.
For both azimuth and elevation, the total angular extent length was defined as the sum of the distances between subsequent local extrema:
\begin{align*}
    TAE = \sum_{k = 0}^{K}|\gamma_{k+1} - \gamma_{k}|
    \centering
\end{align*}
where $\gamma_k$ denotes the orientation at the $k$th local extremum, with $\gamma_0$ and $\gamma_K$ representing the initial and final orientations, respectively, and $K$ is the total number of extrema.

After a preliminary analysis, we decided to exclude frontal positions (azimuth = 0 degrees) from the computation due to the fact that sources were mainly related to reduced movements.

\paragraph{Behavioural data analysis}
Extracted descriptive parameters were analysed at the group level to investigate the effect of the stimulus's orientation and acoustic conditions on task performance.
For each parameter, we computed the average values for each speaker location (M = 27), acoustic condition (L = 4), and participant (N = 12), resulting in 25$\times$4$\times$12$=$1200 values of MO, RT, and Yaw and Tilt TAE.
Since we used a within-subjects paradigm where participants were repeatedly tested, we opted to analyse data by means of Generalized Linear Mixed Models (GLMM) to account for inherent cross-correlation among observations in our dataset and the non-normal data distribution.
Each dependent variable (RT, MO, TAE) was analysed using a generalised linear mixed model (GLMM) with the following structure:
\begin{equation*}
    \text{dep. var.} \sim \text{HRTF} \times (\text{acoustic} + dir) + (1\,|\,\text{ID:$\phi$})
    \centering
\end{equation*}
HRTF and Reverberation were two categorical variables with two levels each (HRTF: individual or non-individual; acoustic: anechoic or reverberant). To address the circular distribution of stimulus positions in the horizontal plane, we defined a categorical factor ($dir$) with three levels corresponding to $90^\circ$ azimuthal sectors: front (azimuth $\leq|45^\circ|$), back (azimuth $\leq|135^\circ|$), and lat (azimuth $\leq90+|45^\circ|$ or $\leq-90+|45^\circ|$).
A random intercept for subjects as a function of stimulus elevation ($\phi$) was included to account for individual differences in baseline task performance.
The interaction terms--HRTF $\times$ reverb (to account for the modulations due to the room acoustic), HRTF $\times$ $dir$ (to account for the modulations due to the horizontal direction)--were introduced to evaluate the modulation of HRTF effect across different reverb conditions and spatial positions.  

Models were fitted using \textit{glmmTMB} \cite{brooks2017glmmtmb} package in R version 4.4.2 (R Core Team, 2020). Specifically, a GLMM with a Gamma family distribution and a logarithmic link function was used (estimated using ML and Nelder-Mead optimizer) to handle the non-linear data distribution of RT \cite{loTransformNotTransform2015}, MO, and Tilt TAE. On the contrary,
Yaw TAE was analysed using an LMM (estimated using REML and nloptwrap optimizer) with a Gaussian family distribution and the identity link, being data better described by a normal distribution.
Model assumptions were validated using the significance of fixed effects, evaluated via Type III Wald chi-square. Post hoc pairwise comparisons for interaction terms were conducted using Tukey-adjusted contrasts (\cite{searlePopulationMarginalMeans1980}).
Model performance was evaluated using marginal and conditional $R^2$, RMSE (root mean squared error), and intra-class correlation coefficient (ICC) using the \textit{performance} package \cite{ludeckePerformancePackageAssessment2021,bonoReportQualityGeneralized2021}. 


\section*{Results}
    \paragraph{Response accuracy}
\mbox{}\\
Participants were able to accurately perform the task as the average percentage of correct responses was high (mean = 98.54$\%$) with low variability across individuals (SD = 1.24$\%$). We report no significant effect of the tested conditions and stimuli's location on the rate of correct responses, indicating that participants were able to perform the task regardless of these factors.

To examine how auditory information influenced task performance in greater detail, we fitted GLMMs on RT, TAE, and MO. Each model included HRTF, acoustic condition (i.e., reverb), and horizontal spatial position as fixed effects, with participant ID and stimulus elevation entered as a random intercept. In terms of performance, models showed a relatively moderate to high explanatory power (conditional $R^2 = [0.43-0.71]$) with modest within-group variability ($ICC = [0.071-0.36]$; see Table~S\ref{supp-tab::table_performance} in the Supplementary Materials for details). Model performance for MO was poor: fixed effects explained modest variance (marginal $R^2 = 0.087$) with a moderate within-group correlation ($ICC = 0.36$). This indicates that most explainable variance lies at the random effects, consistent with large individual variability in MO. Given this limited explanatory power, results for MO are only presented in the Supplementary Information for completeness.

\begin{table}[!h]
\centering
\caption{Combined summary results across multiple models. Estimates and 95\% confidence intervals on the link scale. Statistics are from Type III Wald tests. *p < .05, **p < .01, ***p < .001}
\centering
\resizebox{\ifdim\width>\linewidth\linewidth\else\width\fi}{!}{
\begin{tabular}[t]{llllllll}
\toprule
\multicolumn{2}{c}{\textbf{ }} & \multicolumn{2}{c}{\textbf{RT} [log(sec)]} & \multicolumn{2}{c}{\textbf{TAE Yaw} [deg]} & \multicolumn{2}{c}{\textbf{TAE Tilt} [log(deg)]} \\
\cmidrule(l{3pt}r{3pt}){3-4} \cmidrule(l{3pt}r{3pt}){5-6} \cmidrule(l{3pt}r{3pt}){7-8}
\textbf{Factor} & \textbf{Level} & \textbf{$\chi^2$} & \textbf{Estimate [95\% CI]} & \textbf{$\chi^2$} & \textbf{Estimate [95\% CI]} & \textbf{$\chi^2$} & \textbf{Estimate [95\% CI]}\\
\midrule
\cellcolor{gray!15}{Acoustic} & \cellcolor{gray!15}{} & \cellcolor{gray!15}{12.24***} & \cellcolor{gray!15}{} & \cellcolor{gray!15}{3.17} & \cellcolor{gray!15}{} & \cellcolor{gray!15}{13.76***} & \cellcolor{gray!15}{}\\
\cellcolor{gray!15}{} & \cellcolor{gray!15}{Anechoic} & \cellcolor{gray!15}{} & \cellcolor{gray!15}{0.72 [0.64, 0.79]} & \cellcolor{gray!15}{} & \cellcolor{gray!15}{96.78 [93.70, 99.87]} & \cellcolor{gray!15}{} & \cellcolor{gray!15}{2.54 [2.41, 2.66]}\\
\cellcolor{gray!15}{} & \cellcolor{gray!15}{Reverb} & \cellcolor{gray!15}{} & \cellcolor{gray!15}{0.76 [0.69, 0.83]} & \cellcolor{gray!15}{} & \cellcolor{gray!15}{99.65 [96.56, 102.73]} & \cellcolor{gray!15}{} & \cellcolor{gray!15}{2.59 [2.47, 2.71]}\\
HRTF &  & 12.26*** &  & 0.02 &  & 4.74* & \\
 & Individual &  & 0.71 [0.64, 0.78] &  & 96.40 [93.30, 99.50] &  & 2.60 [2.47, 2.72]\\
 & KEMAR &  & 0.76 [0.69, 0.83] &  & 100.03 [96.93, 103.13] &  & 2.53 [2.41, 2.65]\\

 \cellcolor{gray!15}{dir} & \cellcolor{gray!15}{} & \cellcolor{gray!15}{332.86***} & \cellcolor{gray!15}{} & \cellcolor{gray!15}{1540.99***} & \cellcolor{gray!15}{} & \cellcolor{gray!15}{218.40***} & \cellcolor{gray!15}{}\\
\cellcolor{gray!15}{} & \cellcolor{gray!15}{back} & \cellcolor{gray!15}{} & \cellcolor{gray!15}{1.02 [0.94, 1.09]} & \cellcolor{gray!15}{} & \cellcolor{gray!15}{150.43 [147.13, 153.73]} & \cellcolor{gray!15}{} & \cellcolor{gray!15}{3.00 [2.87, 3.12]}\\
\cellcolor{gray!15}{} & \cellcolor{gray!15}{front} & \cellcolor{gray!15}{} & \cellcolor{gray!15}{0.55 [0.47, 0.62]} & \cellcolor{gray!15}{} & \cellcolor{gray!15}{56.10 [52.42, 59.79]} & \cellcolor{gray!15}{} & \cellcolor{gray!15}{2.45 [2.32, 2.58]}\\
\cellcolor{gray!15}{} & \cellcolor{gray!15}{lat} & \cellcolor{gray!15}{} & \cellcolor{gray!15}{0.64 [0.57, 0.72]} & \cellcolor{gray!15}{} & \cellcolor{gray!15}{88.11 [84.83, 91.40]} & \cellcolor{gray!15}{} & \cellcolor{gray!15}{2.24 [2.11, 2.36]}\\
HRTF × dir &  & 7.47* &  & 9.13* &  & 6.94* & \\
 & Individual:back &  & 0.98 [0.91, 1.06] &  & 150.59 [146.55, 154.62] &  & 2.99 [2.85, 3.12]\\
 & Individual:front &  & 0.50 [0.42, 0.59] &  & 50.95 [46.30, 55.61] &  & 2.55 [2.40, 2.69]\\
 & Individual:lat &  & 0.65 [0.57, 0.73] &  & 87.65 [83.69, 91.61] &  & 2.26 [2.12, 2.39]\\
\addlinespace
 & KEMAR:back &  & 1.05 [0.98, 1.13] &  & 150.26 [146.23, 154.30] &  & 3.01 [2.87, 3.14]\\
 & KEMAR:front &  & 0.59 [0.51, 0.67] &  & 61.25 [56.60, 65.91] &  & 2.36 [2.22, 2.50]\\
 & KEMAR:lat &  & 0.64 [0.56, 0.72] &  & 88.58 [84.62, 92.53] &  & 2.22 [2.08, 2.35]\\
 \cellcolor{gray!15}{HRTF × Acoustic} & \cellcolor{gray!15}{} & \cellcolor{gray!15}{4.53*} & \cellcolor{gray!15}{} & \cellcolor{gray!15}{0.25} & \cellcolor{gray!15}{} & \cellcolor{gray!15}{12.54***} & \cellcolor{gray!15}{}\\
\cellcolor{gray!15}{} & \cellcolor{gray!15}{Individual:Anechoic} & \cellcolor{gray!15}{} & \cellcolor{gray!15}{0.67 [0.60, 0.75]} & \cellcolor{gray!15}{} & \cellcolor{gray!15}{94.61 [90.93, 98.28]} & \cellcolor{gray!15}{} & \cellcolor{gray!15}{2.52 [2.39, 2.65]}\\
\cellcolor{gray!15}{} & \cellcolor{gray!15}{Individual:Reverb} & \cellcolor{gray!15}{} & \cellcolor{gray!15}{0.75 [0.67, 0.83]} & \cellcolor{gray!15}{} & \cellcolor{gray!15}{98.19 [94.51, 101.86]} & \cellcolor{gray!15}{} & \cellcolor{gray!15}{2.68 [2.55, 2.81]}\\
\cellcolor{gray!15}{} & \cellcolor{gray!15}{KEMAR:Anechoic} & \cellcolor{gray!15}{} & \cellcolor{gray!15}{0.76 [0.68, 0.83]} & \cellcolor{gray!15}{} & \cellcolor{gray!15}{98.95 [95.28, 102.63]} & \cellcolor{gray!15}{} & \cellcolor{gray!15}{2.56 [2.43, 2.68]}\\
\cellcolor{gray!15}{} & \cellcolor{gray!15}{KEMAR:Reverb} & \cellcolor{gray!15}{} & \cellcolor{gray!15}{0.77 [0.69, 0.84]} & \cellcolor{gray!15}{} & \cellcolor{gray!15}{101.11 [97.43, 104.78]} & \cellcolor{gray!15}{} & \cellcolor{gray!15}{2.50 [2.37, 2.63]}\\
\bottomrule
\end{tabular}}
\label{tab::model_results_tot}
\end{table}

\begin{figure}[hbt!]
    \centering
    \includegraphics{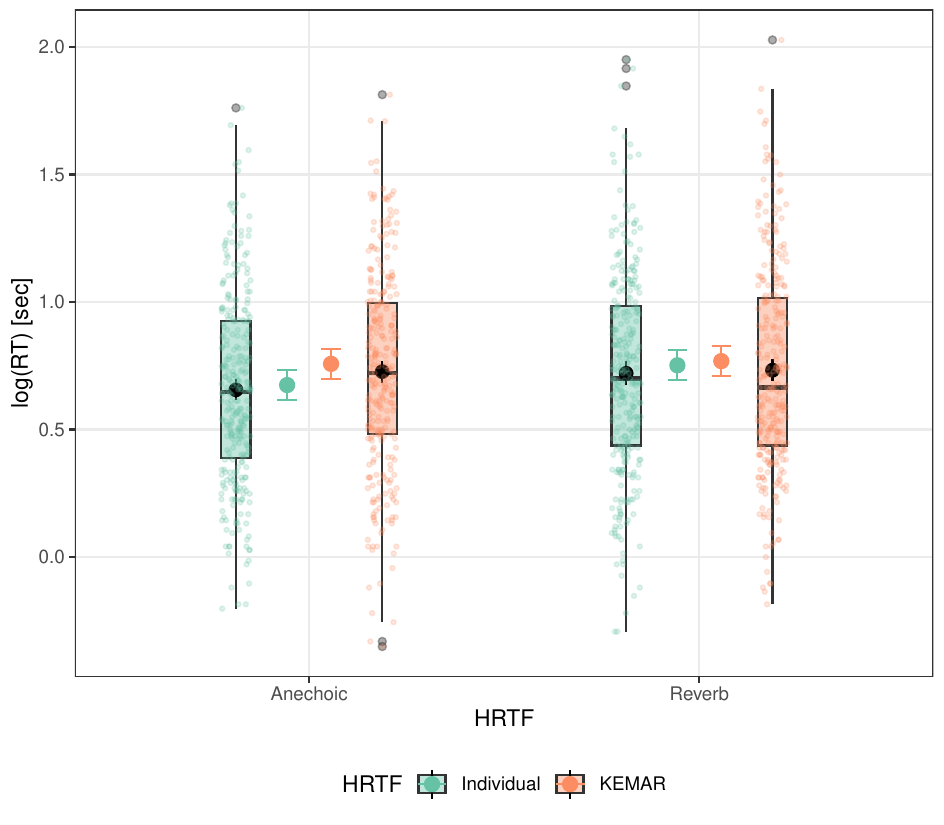}
    \caption{\textbf{Response time as a function of acoustic and HRTF: data distribution and model estimates}. Interaction of HRTF (green: individual, orange: KEMAR) and acoustic on response time (\textit{x-axis}). Dots show estimated response time, and the interval shows the 95\% confidence interval of the estimate. Data are shown in link scale for clearer comparison. 
    }
    \label{fig:RTVsCond}
\end{figure}

\paragraph{Response time analysis}
\mbox{}\\
Because RT was the primary outcome, we first report its main effects and interactions. Analyses of head strategies and TAE (Yaw and Tilt direction) are presented as exploratory to assess whether similar patterns extend to listening strategy.
A comprehensive summary of the analysis for all investigated fixed effects, along with their relative statistical significance, is provided in Table \ref{tab::model_results_tot}.
The GLMM revealed significant main effects of HRTF ($\chi^2(1) = 12.26$, $p < 0.001$) and acoustic ($\chi^2(1) = 12.24$, $p < 0.001$). 
The model also yielded a significant interaction ($\chi^2(1) = 4.53$, $p = 0.03$) indicating that the HRTF effect changed across acoustic conditions, even though the observed effect size was qualitatively small (see Figure~\ref{fig:RTVsCond}). 
Simple effect analysis showed that response times were significantly faster with individual HRTFs in the anechoic condition (individual: $0.67$\,log(s), 95\% CI $[0.60, 0.75]$; non-individual: $0.76$\,log(s), $[0.68, 0.83]$; $z = -3.72$, $p = 0.0002$), but not in the reverberant condition (individual: $0.75$\,log(s), $[0.60, 0.75]$; non-individual: $0.77$\,log(s), $[0.69, 0.84]$; $z = -0.745$, $p = 0.456$). 
This pattern suggests that room reflections reduced or masked the benefit provided by individualized pinna cues.

\begin{figure}[hbt!]
    \centering
    \includegraphics{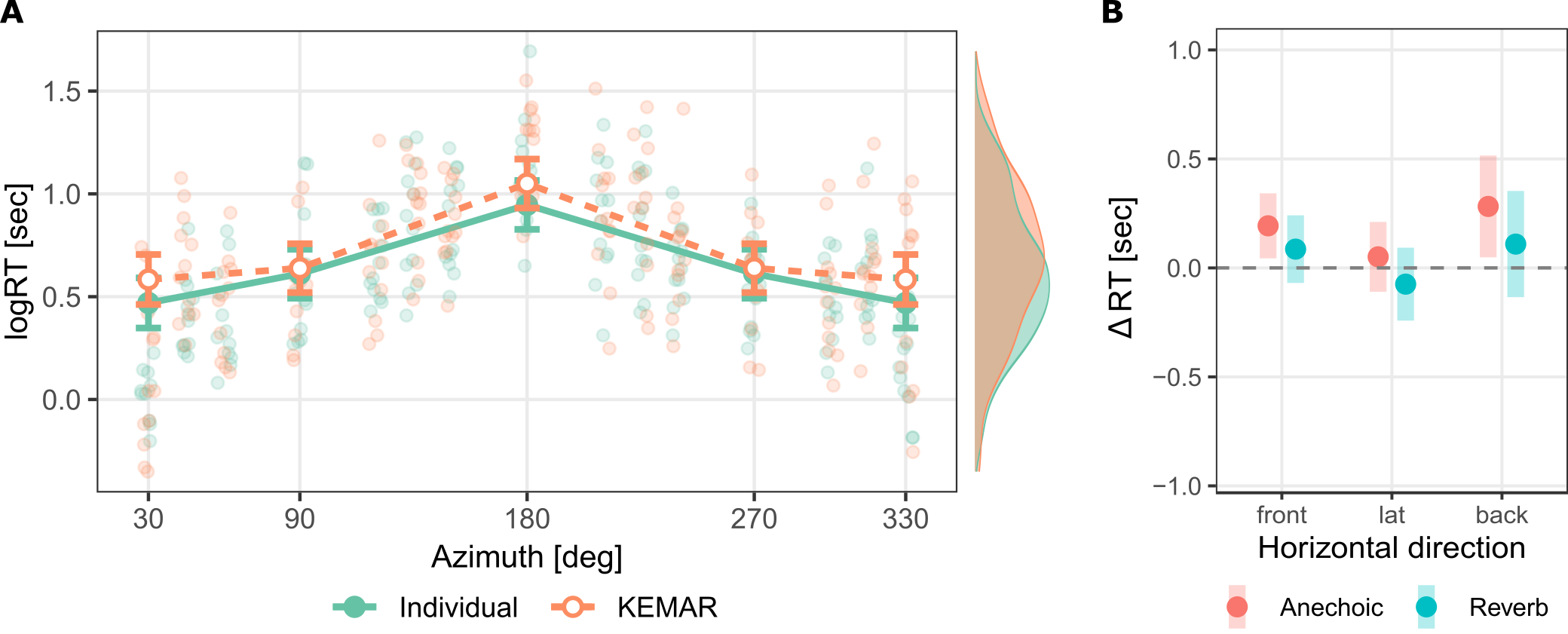}
    \caption{\textbf{A.} Response time in the anechoic condition as a function of azimuth (degrees) and HRTF (green indicates individual, orange indicates KEMAR). Scatter points show the raw data, and the data distributions are shown on the right. Model estimates are represented by solid lines, and the intervals show the 95\% confidence interval. Data are shown in the link scale for clearer comparison.
    \textbf{B.} Difference in estimated response time between individual and KEMAR HRTFs as a function of horizontal direction and acoustic condition (red indicates anechoic, blue indicates reverb).}
    \label{fig:RTvsDir}
\end{figure}

Since the direction of the stimulus likely influenced the relative advantages of HRTF individualization, we examined the interaction between HRTF and the horizontal target direction (dir) in the anechoic condition. GLMM showed a significant difference ($\chi^2(1) = 7.47$, $ p = 0.03$), which was confirmed by the simple effect analysis on the HRTF factor as a function of dir. Significant effects were found for back ($z = -3.50$, $p = 0.0005$) and front ($z = -3.27$, $p = 0.0011$) directions, but not for lateral ones ($z = -0.95$, $p = 0.342$) (see Figure ~\ref{fig:RTvsDir}A)
This specific directional dependency may be related to the different front-back confusion rates when the two different HRTFs were used. A summary of this analysis is shown in Figure~\ref{fig:RTvsDir}B.

\begin{figure}[!h]
    \centering
    \includegraphics{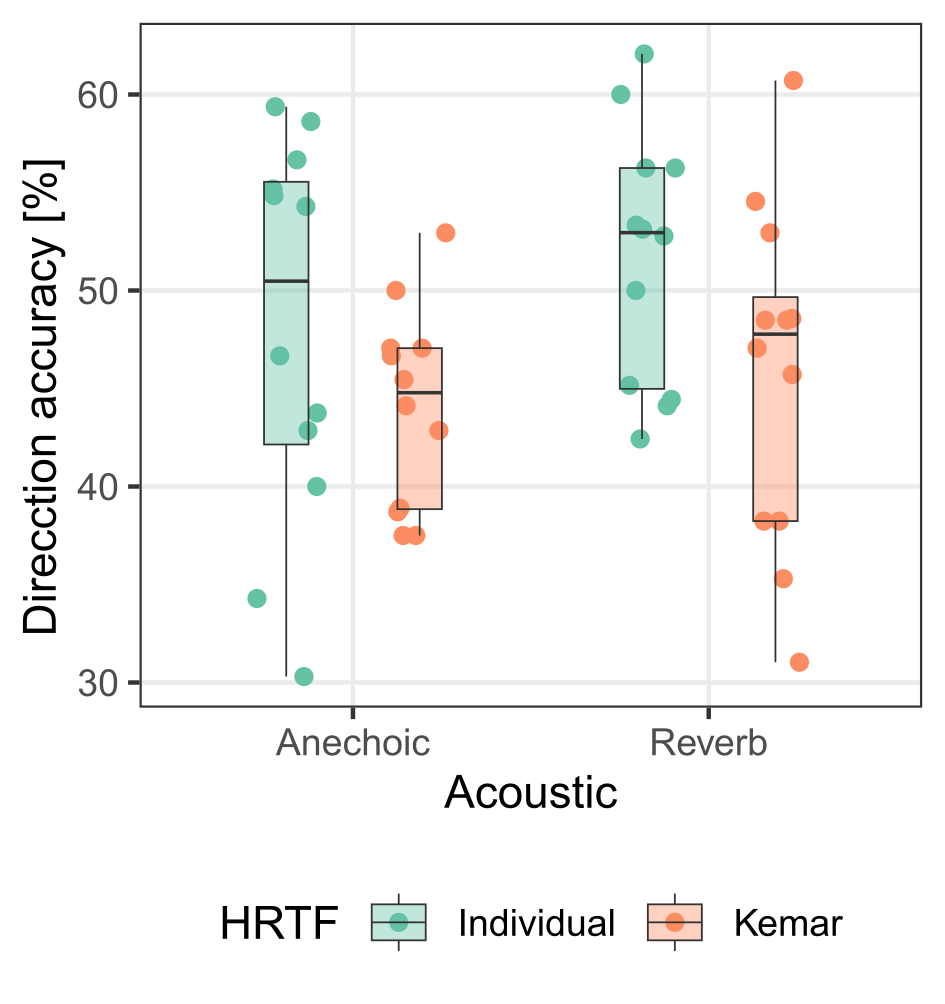}
    \caption{\textbf{Estimated initial directional accuracy in up-down discrimination.} Directional accuracy distribution at movement onset in up-down discrimination as a function of acoustic and HRTF (green: individual; orange: KEMAR). Points represent estimated individual accuracy.}
    \label{fig:up-down}
\end{figure}

\begin{figure}[!h]
    \centering
    \includegraphics[width = \textwidth, keepaspectratio]{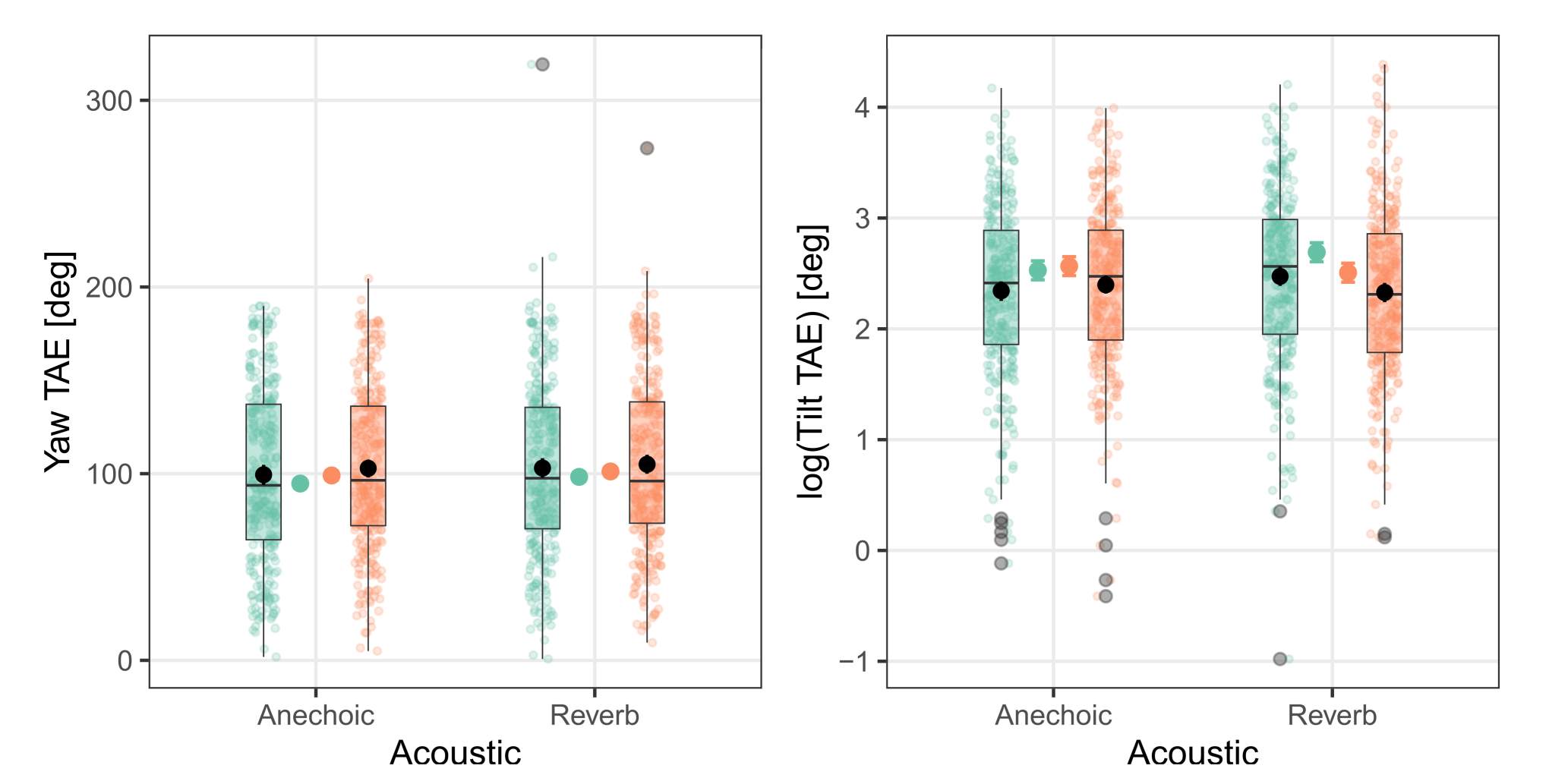}
    \caption{\textbf{Effect of HRTF and acoustic on movement extents} \textbf{A.} Interaction of Acoustic (\textit{x-axis}) and HRTF (green: individual, orange: non-individual) on Yaw (left) and Tilt (right) Total Angular Extent (TAE). Values are shown in the link scale for easier comparison. Coloured Dots show estimated marginal means; intervals indicate 95\% confidence intervals. Data distribution: black dots (mean), black line (median).}
    \label{fig:parVsCond_effects}
\end{figure}

\begin{figure}[hbt!]
    \centering
    \includegraphics{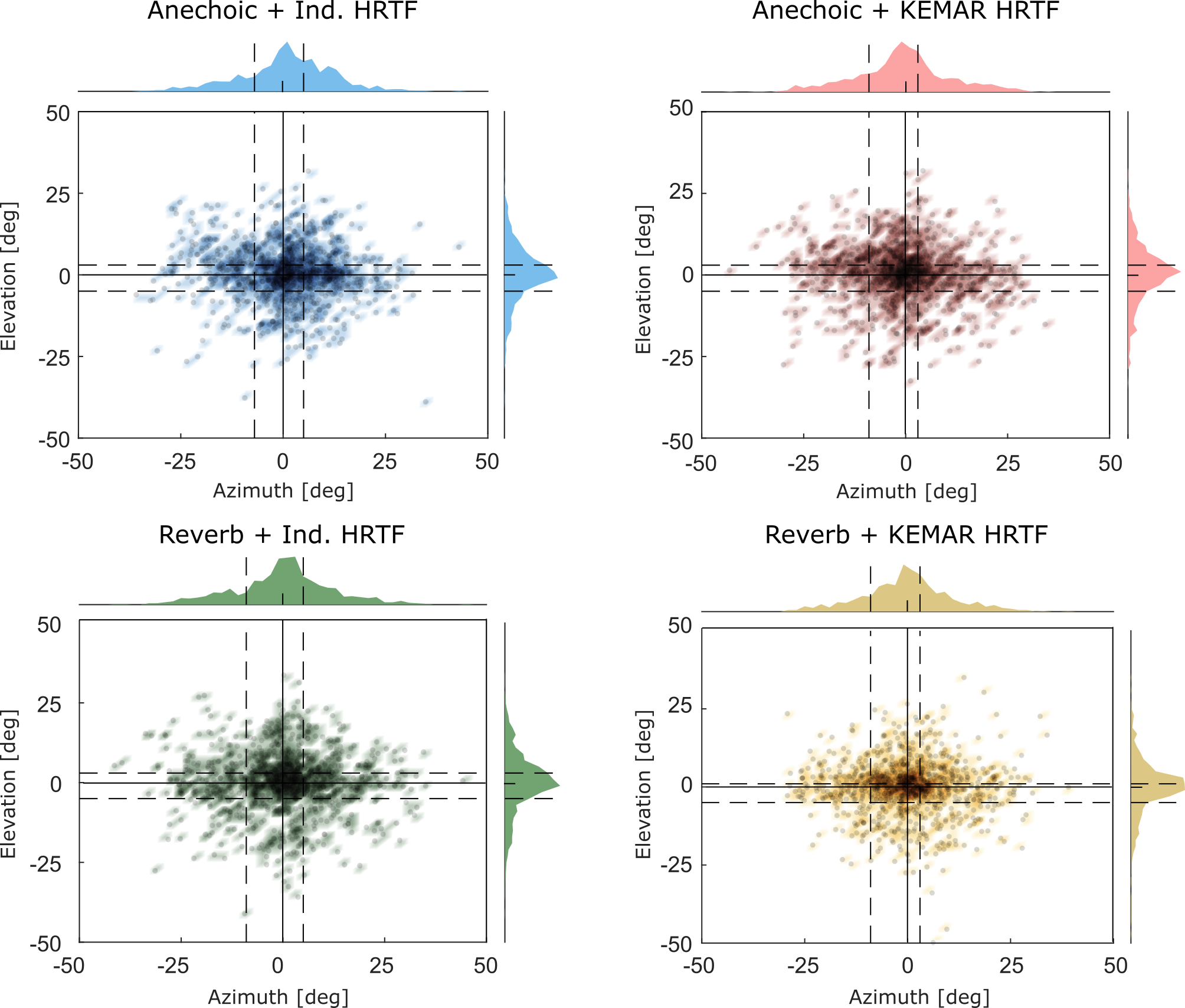}
    \caption{\textbf{Angular head misalignment relative to target location in tested conditions.}
    Deviation (in degrees) between the final head orientation and the actual target position as a function of tested acoustic conditions. The marginal distribution of errors illustrates the frequency of misalignment across the angular axis.
    }
    \label{fig:generic-behavior}
\end{figure}

\paragraph{Analysis of head-rotation behaviour}
\mbox{}\\
To assess whether differences in search time across conditions were associated with changes in head-movement behaviour, we examined (i) the extent of exploration and (ii) the execution of head movements.\\
\textit{Exploration extent.}
Figure~\ref{fig:parVsCond_effects}A–B shows yaw and tilt TAE as a function of HRTF and acoustic condition. In the yaw direction, no significant main effect of HRTF or acoustic condition was observed, and their interaction was not significant (Table~\ref{tab::model_results_tot}), indicating comparable horizontal exploration when averaged across stimulus directions. When stimulus direction was included, a significant interaction between HRTF and direction was found (\textbf{$p = 0.03$}). Post-hoc analyses revealed increased yaw TAE for frontal stimuli with the KEMAR HRTF, indicating greater horizontal exploration for this subset of trials, likely due to increased front-back confusion.
In the tilt direction, significant effects of HRTF ($\chi^2(1) = 4.74$, $p = 0.028$), acoustic condition ($\chi^2(1) = 13.76$, $p < 0.001$), and their interaction ($\chi^2(1) = 12.54$, $p < 0.001$) were observed. Simple-effect analyses showed a small increase in vertical exploration with the individual HRTF in the reverberant condition (Table~\ref{tab::model_results_tot}). Tilt movements were more variable across participants than yaw movements.\\
\textit{Head orientation planning}. 
To better understand the relative influence of acoustic condition on initial motor planning, we analysed the movement direction at onset relative to the stimulus location direction. The deviation between the initial movement direction and the stimulus direction was used to estimate up–down directional confusion. Figure \ref{fig:up-down} shows the distribution of subjective estimated initial directional accuracy values, computed for each participant as a function of acoustic condition.
Repeated measure anova revealed a significant effect for HRTF ($\chi^2(1) = 5.9363, p = 0.01483$) with a large effect size (partial $\eta = 0.15$), suggesting a significant improvement of percentage of correctness in the direction of the movement of movement initiation when an individual HRTF was used.
No significant effect were reported neither for acoustic condition ($\chi^2(1) = 1.7090, p = 0.19112$; partial $\eta = 0.05$) or the interaction term ($\chi^2(1) = 0.2146, p = 0.64319$; partial $\eta = 0.00646$).\\
\textit{Head–response alignment.}
Finally, we assessed the alignment between head gaze direction and final response location. 
By evaluating head alignment errors relative to target directions (see Figure~\ref{fig:generic-behavior}), we observed that while head positioning was accurate on average (azimuth: $mean = -0.018^\circ$, elevation: $mean = 0.29^\circ$), it lacked precision, exhibiting substantial variability in both horizontal (azimuth: $SD = 15.58$$^\circ$) and vertical (elevation: $SD = 11.98^\circ$) directions. Given the high response accuracy (98.54\%), this low precision confirms that participants utilized head movements primarily for coarse repositioning, relying on compensatory eye movements to refine the visual search and achieve the final target selection.


\section*{Discussion and conclusions}
    \paragraph{Summary}
In this study, we examined how HRTF individualisation affects performance and movement organisation during an aurally guided visual search task under anechoic and reverberant conditions. Rather than treating localisation as a purely perceptual outcome, we approached it as a sensorimotor process in which auditory spatial estimates are immediately translated into orienting actions.

Under anechoic conditions, individualised HRTFs led to faster responses than non-individualised HRTFs ($\Delta RT = 200$ ms), particularly for front--back stimuli. This effect was modest in magnitude but statistically significant and was expressed primarily in movement initiation, whereas overall movement extent was only slightly affected. These findings suggest that the benefit of individualisation in dynamic tasks is subtle and context-dependent.
In reverberant conditions, this advantage disappeared. The absence of HRTF-dependent differences suggests that the benefit of individualisation depends on the availability of reliable spectral cues. When reverberation reduces the reliability of these cues, the advantage observed during early orienting is no longer apparent.
Taken together, the results point to a subtle but consistent pattern that HRTF individualisation appears to influence sensorimotor organisation in spatial hearing.

\paragraph*{VR paradigm validation: data comparison}

Our median RT (1.98$~$s, SD$~=~$1.04$~$s) is comparable to values reported in previous studies using VR-based binaural paradigms \cite{boliaAurallyAidedVisual1999, cunioSpatialAuditoryCueing2019}, but exceeds those observed in free-field paradigms. In such paradigms where multi-loudspeaker systems are coupled with LEDs for visual stimulation, RTs of approximately 1 s have been reported for listeners with unoccluded ears (see \cite{simpsonImpactHearingProtection2005,lladoImpactHeadwornDevices2024}).

This difference is reasonable because the combination of VR and binaural reproduction introduces additional perceptual and computational demands. Listeners must interpret simulated auditory and visual cues that do not fully reproduce real-world acoustics and visual context \cite{cummingsHowImmersiveEnough2016}. Even when individualised, virtual spatial audio does not perfectly replicate free-field stimulation, in which front--back confusions are minimal and localisation performance can approach ceiling levels \cite{middlebrooksVirtualLocalizationImproved1999}.

Moreover, the longer RTs observed in our study may also be related to the shorter familiarisation phase. Unlike the aforementioned studies, we intentionally omitted training to minimise adaptation to the HRTFs, as our aim was to investigate spatial-hearing performance before substantial perceptual adjustment could occur. Visual and proprioceptive feedback, including that provided through head tracking, have been shown to support adaptation to HRTFs and can lead to localisation performance comparable to that obtained with poorly matched HRTFs \cite{steadmanShorttermEffectsSound2019,picinaliSystemtoUserUsertoSystemAdaptations2023}. Excluding training therefore allowed us to characterise initial, relatively unadapted spatial-hearing performance rather than compensatory mechanisms acquired through perceptual learning.

\paragraph*{Non-Individual HRTFs led to increased response times in anechoic conditions}

Previous research has consistently demonstrated that individualised HRTFs improve localisation accuracy in static, anechoic conditions compared with non-individualised HRTFs \cite{liMeasurementHeadRelatedTransfer2020,guezenocHRTFIndividualizationSurvey2018}. In dynamic contexts, however, the available evidence remains limited and partly inconsistent \cite{rummukainenHeadRelatedTransferFunctions2021,oberemExperimentsLocalizationAccuracy2020,ben-hurLocalizationVirtualSounds2020,jennyUsabilityIndividualizedHeadRelated2020,begaultDirectComparisonImpact2001}. When listeners are allowed to move, dynamic cues can reduce front--back confusions even with non-individualised HRTFs.
The present results support an intermediate interpretation. Under anechoic conditions, individualised HRTFs produced a measurable, although modest, reduction in response time, particularly for front--back stimuli. Crucially, this advantage was expressed primarily in movement initiation rather than in total angular extent or later task performance. In other words, individualisation appeared to facilitate the decision to commit to an initial orienting direction, while having a more limited effect once the movement had unfolded.
This pattern is consistent with the possibility that dynamic head movements do not eliminate the benefit of individualisation, but instead compensate for early uncertainty over time. Non-individualised HRTFs may increase ambiguity in the initial spatial estimate, particularly along the front--back axis, thereby delaying action onset. Once the movement sequence begins, however, dynamic sampling and emerging visual feedback can reduce these initial differences, resulting in more similar behavioural outcomes across conditions.

In reverberant conditions, the initiation advantage disappeared. The absence of significant HRTF-dependent differences suggests that early reflections and temporal smearing may have reduced the perceptual salience of individualised spectral cues, thereby increasing reliance on dynamic information and environmental acoustics. This interpretation is consistent with findings from \cite{begaultAuditoryNonAuditoryFactors2001}, who reported no benefit of individualisation during active localisation in reverberant environments, as well as with more recent work showing minimal HRTF effects on speech-based tasks under reverberation \cite{daugintisEffectsBinauralRendering2024,orduna-bustamanteBinauralSpeechIntelligibility2018}.

Together, these results suggest that individualisation matters most when spectral cues are both reliable and behaviourally relevant for spatial disambiguation. When reverberation reduces their reliability, listeners may rely more strongly on dynamic sampling strategies.

\paragraph*{An interpretation of aurally-guided visual search as a sensorimotor process}

Sensorimotor theory proposes that perception is an active process shaped by learned relationships between action and sensory change \cite{o2001sensorimotor,gibsonEcologicalApproachVisual2014,froese2020passive}. Within this framework, auditory localisation can be interpreted as an action-guiding process in which an initial spatial estimate is continuously refined through movement \cite{aytekin2008sensorimotor}.

Ege et al. (2018) \cite{egeAccuracyPrecisionTradeoffHuman2018a} showed that horizontal and vertical orienting rely on partially distinct control dynamics shaped by cue reliability. Azimuth is supported by robust interaural differences that enable relatively stable directional movements. In contrast, elevation depends more strongly on pinna-mediated spectral cues, which are more susceptible to degradation and are therefore associated with greater variability during movement planning. Such differences suggest that the auditory system can adapt the organisation of movement according to the reliability of the spatial cues available. From this perspective, HRTF individualisation may be relevant primarily because it influences the reliability of the initial spatial estimate that guides action. Its effect may therefore be reflected more strongly in how efficiently the listener begins to move towards a target than in where the listener ultimately arrives.

Our findings are consistent with this interpretation. First, the fastest and most accurate movement initiations were observed for lateral stimuli, for which robust interaural differences provide a reliable directional signal largely independently of HRTF condition. For frontal and rear positions, by contrast, binaural cues are more ambiguous and spectral information becomes particularly important for disambiguating the initial direction of movement.
Second, azimuthal planning was relatively stable across conditions, whereas vertical planning was more sensitive to HRTF condition, reflecting the different reliability of binaural and monaural spectral cues. Third, the individualisation advantage in response time disappeared under reverberant conditions, in which spectral structure was more strongly degraded. Together, these observations support the interpretation that the behavioural benefit of individualisation is most apparent when individualised spectral cues remain informative.
Importantly, movement planning was sensitive to HRTF condition but not to reverberation, whereas response time was affected by both. This pattern hints at a possible dissociation between early motor commitment and overall response organisation, suggesting that reverberation and HRTF mismatch may influence partially different stages of the sensorimotor process. However, the present data do not allow this possibility to be interpreted conclusively, and it warrants targeted investigation in future studies.

Under a sensorimotor account, changes in acoustic cue reliability might also be expected to influence the organisation of exploratory behaviour. We suggest that head movements alone may not be the primary mechanism affected. Instead, the coordination between eye and head movements may be particularly relevant. During natural orienting, saccadic eye movements typically precede and guide slower head rotations, while efficient gaze anchoring depends on a sufficiently reliable initial spatial estimate \cite{goossensHumanEyeheadCoordination1997,vliegenDynamicSoundLocalization2004}.
If individualised HRTFs improve the reliability of the estimate that initiates this sequence, they may facilitate faster gaze stabilisation and earlier head movement onset, consistent with the movement-initiation effects observed here. With non-individualised HRTFs, the initial orienting response may instead be delayed. Once the orienting sequence is underway, however, visual and multisensory feedback may compensate for this initial uncertainty, potentially explaining why later task outcomes remain relatively preserved. Because eye movements were not measured in the present study, this mechanism should be regarded as a testable prediction rather than an established result.
Taken together, the observed pattern is consistent with a sensorimotor reliability account in which HRTF individualisation modulates the confidence with which the auditory system commits to an orienting action. This effect appears most clearly during the early stages of movement, particularly for spatial dimensions in which spectral cues play an important role and under acoustic conditions in which those cues remain informative.

\paragraph*{Limitations}

If active-perception frameworks predict that changes in cue reliability should reorganise exploratory strategies, the relatively modest magnitude of the observed effects warrants consideration. Several methodological constraints may have attenuated the behavioural expression of the HRTF manipulation.

Real-time 6DoF (six-degree-of-freedom) rendering may have reduced the contrast between individualised and non-individualised HRTFs. Continuous movement requires HRTF interpolation, which can introduce magnitude errors, particularly at high frequencies and in contralateral regions \cite{arend2023magnitude}. These interpolation artefacts may therefore have weakened the individualisation effect observed in our experiment.
A second limitation concerns the reverberant condition specifically. In the present setup, individualisation was applied only to the direct sound path using the 3DTI framework \cite{cuevas-rodriguez3DTuneInToolkit2019}, whereas the reverberant field was rendered using a non-individualised BRIR (Binaural room impulse response) derived from a KEMAR mannequin. Consequently, the absence of an individualisation effect under reverberation may show this limitation rather an ecologically valid effect. Fully individualised BRIRs, in which both direct and reflected components incorporate listener-specific filtering, would provide a cleaner test of whether the reliability advantage of individualisation persists under room-acoustic conditions \cite{neidhardt2022perceptual}.
Finally, although participants were free to move, the experimental context remained more constrained than natural listening environments. Ecological approaches to spatial hearing emphasise not only stimulus complexity but also the degree of agency and exploratory freedom available to the listener \cite{o2001sensorimotor,gibsonEcologicalApproachVisual2014}. Subtle constraints imposed by the available movement range or task structure may have limited the extent to which participants deployed active perceptual strategies, thereby reducing observable differences between conditions.

\paragraph*{Conclusion and future directions. Towards realistic listening tests}

The present findings show that HRTF individualisation can affect the timing and organisation of orienting behaviour even when later task outcomes remain similar. This suggests that dynamic listening tests should consider movement-related measures alongside conventional localisation metrics.
Future paradigms should allow greater freedom of movement and richer acoustic interaction while maintaining experimental control. Eye tracking could further reveal whether HRTF mismatch affects the temporal coordination of gaze and head orientation \cite{vliegenDynamicSoundLocalization2004,goossensHumanEyeheadCoordination1997}.
More accurate spatial rendering will also be important, particularly in reverberant conditions. Fully individualised rendering of both direct and reflected sound would provide a cleaner test of whether the effects observed here persist in more realistic acoustic environments.


\bibliography{biblio}


\section*{Data availability}
The data generated and used in this study will be made available from the corresponding author upon request.

\section*{Acknowledgements}
Claude (Anthropic) was used as a writing assistant for language editing and clarity. All scientific content, interpretations, and conclusions were reviewed and approved by the authors, who bear full responsibility for the final text.

This manuscript represents a first presentation of these data and is intended as a preliminary version of the work.

\section*{Author information}

\subsection*{Authors and affiliations}
\textbf{Department of Informatics, Bioengineering, Robotics and Systems Engineering, University of Genoa, Genoa, Italy}
\textbf{Dyson School of
Design Engineering, Imperial College London, London, United Kingdom}
\newline
Fulvio Missoni, Katarina C. Poole, Tim Murray-Browne, Andrea Canessa \& Lorenzo Picinali

\subsection*{Contributions}
K.C.P. conceived and designed the experimental paradigm. T.M.B. contributed to paradigm design and led the experimental implementation and software development. F.M. contributed to the experimental implementation, conducted HRTF measurements and data acquisition, performed the data analysis in MatLab and R, interpreted the results, and drafted the manuscript. L.P. contributed to the discussion and interpretation of the results and provided critical revision of the initial manuscript. A.C. contributed to the discussion and interpretation of the results and reviewed the manuscript.

\subsection*{Corresponding author}
Correspondence to Fulvio Missoni at fulvio.missoni@edu.unige.it.

\subsection*{Competing interests}
The authors declare no competing interests.

\end{document}


\begin{table}[h!]
  \centering
      \caption{\textbf{Model structure and relative performance} For each model, the used family distribution and the relative link function are presented. Model performance parameters are the following.
    $R^2$ (conditional): r-squared value conditional, considering both fixed and random effects. $R^2$ (marginal): r-squared value marginal, considering only fixed effects of the model. ICC: intra-class correlation coefficient. RMSE: root mean squared error expressed in the estimator's units. AIC: Akaike's Information Criterion. BIC: Bayesian Information Criterion.}
  \begin{adjustbox}{width=\textwidth}
    {\tabulinesep=1.2mm
   \begin{tabu}{lcccc}
      \toprule
      & RT & Yaw Total Angular Extent & Tilt Total Angular Extent & MO \\
      \midrule
        Family Dist.                        & Gamma     &  Gaussian & Gamma    & Gamma      \\
        Link Function                       & Log       & Identity  & Log      & Log        \\
        R\textsuperscript{2} (conditional)  &    0.56   &     0.71  &     0.50 &     0.42   \\
        R\textsuperscript{2} (marginal)     &    0.31   &     0.69  &     0.35 &     0.09   \\
        ICC                                 &    0.36   &     0.07  &     0.23 &     0.36   \\
        RMSE                                &    0.66   &    24.46  &     8.73 &   198.72   \\
        AIC                                 & 2017.81   & 11139.61  &  7801.25 & 32173.19   \\
        BIC                                 & 2078.89   & 11139.35  &  7862.34 & 32234.27   \\
      \bottomrule
    \end{tabu}}
  \end{adjustbox}
  \label{tab::table_performance}
\end{table}